\documentclass{article}
\usepackage{spconf,amsmath,graphicx,hyperref, booktabs, makecell, multirow, amssymb, cite, enumitem}

\title{NECTAR: Automatic Negative Clue Estimation via \\ Centroid-Space Mapping for Target Speaker Extraction}

\name{Hyungjun Yoo$^{1}$ \qquad Yongjoon Lee$^{1}$ \qquad Jung-Woo Choi$^{2 *}$\thanks{$^*$Corresponding author.}}
\address{$^{1}$School of Electrical Engineering, KAIST, Daejeon, South Korea \\
$^{2}$Department of AI Systems, KAIST, Daejeon, South Korea}

\begin{document}
\ninept
\maketitle

\begin{abstract}
Target speaker extraction (TSE) aims to extract a desired speaker from a mixture using auxiliary clues such as enrollment speech. 
An interferer-related representation can provide a useful negative clue for TSE, but such information is not directly available from the target enrollment.
To derive this representation from the observed mixture, we analyze the relationship between the speaker representations of a speech mixture and its constituent speakers.
Our analysis shows that the mixture centroid lies near the midpoint of its constituent speaker centroids, whereas this relationship does not hold for single-utterance embeddings.
Based on this observation, we propose NECTAR, NEgative Clue for TARget speaker extraction, which automatically derives an interferer-related negative clue from the mixture using the centroid-space relationship.
We further introduce a Centroid-Space Mapper (CSM) to estimate the required centroid representations from a mixture and a target enrollment.
NECTAR achieves the highest SI-SDRi on Libri2Mix among the compared systems trained on LibriSpeech train-clean-100, reaching 16.18~dB. 

\end{abstract}


\section{Introduction}
\label{sec:intro}

Target speaker extraction (TSE) aims to extract a desired speaker from a speech mixture using auxiliary clues that identify the target. In particular, target enrollment speech is widely used as a TSE clue. Embedding-free approaches directly exploit enrollment signals~\cite{direct_context_tse,sefnet,usef_tse}, whereas embedding-based approaches encode the
enrollment into a latent speaker representation to condition the
system~\cite{hierarchical,multilevel_tse,ssl_tse,
selective_hubert,aug4tse}. Within embedding-based TSE, considerable effort has focused on improving the target-speaker representation.
Recent studies have explored hierarchical~\cite{hierarchical} or multi-level speaker
representations~\cite{multilevel_tse}, 
self-supervised speech representations~\cite{ssl_tse,selective_hubert}, and the augmentation of short enrollment speech~\cite{aug4tse}.
Despite these advances, most conventional embedding-based systems primarily condition the extractor on the target speaker.

Recent work has shown that treating an interfering speaker’s enrollment as a negative cue to suppress that speaker in the mixture is an effective strategy for TSE~\cite{chime9_tse}.
This raises a key question: \textit{Can interferer-related information, or a negative clue, be derived without additional interferer enrollment?}
To address this question, we analyze speaker-embedding geometry in centroid space and establish the relation between the mixture centroid and the centroids of its constituent speakers.
This geometric relationship provides a way to estimate an interferer-related representation from the mixture and target-speaker representations, enabling negative-clue conditioning without explicit interferer enrollment.

Building on this structure, we propose \textbf{NE}gative \textbf{C}lue for \textbf{TAR}get speaker extraction \textbf{(NECTAR)}, a TSE framework that automatically derives an interferer-related negative clue from the mixture and leverages both target and interferer information.
Since the centroid-space relationship does not directly hold for individual utterance embeddings, we introduce a Centroid-Space Mapper (CSM) to estimate centroid-space representations from individual input segments, allowing the relationship to be exploited for negative clue estimation. The code is available at the following link\footnote{\url{https://github.com/Hyungjun-Yoo811/NECTAR}}. The main contributions of this work are as follows:
\begin{itemize} [leftmargin=1.2em,itemsep=1pt,topsep=2pt,parsep=0pt]

    \item We discover a \textbf{Centroid-Space Relationship} among the target, interferer, and mixture centroids, where the mixture centroid lies
    near the midpoint of the speaker centroids.
    
    \item We propose \textbf{NECTAR}, an embedding-based TSE framework that automatically derives a negative clue from the mixture and jointly uses target and interferer information.

    \item Our method is effective across in-domain, out-of-domain, and noisy settings, with negligible computational overhead. It achieves an SI-SDRi of 16.18 dB and an extraction accuracy of 97.1\% on Libri2Mix, outperforming all compared systems.
    

\end{itemize}

\section{Related Work}
\label{sec:related_work}

\subsection{Speaker Centroid Estimation}

Speaker centroids aggregate multiple utterance embeddings into speaker-level representations.
GE2E~\cite{ge2e} constructs centroids from multiple utterances for speaker-discriminative embedding learning, while metric-learning
approaches~\cite{metric_learning} similarly exploit speaker-level aggregation to improve embedding discrimination.
For TSE, Heo et al.~\cite{centroid_tse} estimated a speaker centroid from a single enrollment utterance, providing a more robust target-speaker representation.
These studies used speaker centroids to improve speaker representations, whereas our work focuses on the geometric relationship between mixture and speaker centroids to derive a negative clue.

\subsection{Negative Clues for Source Extraction}

In target sound extraction, query-conditioned methods have enabled
selective extraction or removal using class-based queries~\cite{sound_selector}, positive/negative queries~\cite{clapsep}, and multimodal queries~\cite{omnisep,promptsep}.
Speaker embedding de-mixing~\cite{speaker_demix} reconstructs the other speaker embedding from a mixture and a known speaker embedding, while Zhao et al.~\cite{chime9_tse} used interfering-speaker enrollment signals as a negative clue. In contrast, NECTAR exploits an explicit geometric relationship between mixture and speaker centroids and derives a negative clue, without requiring interferer enrollment at inference.

\section{NECTAR}
\label{sec:nectar_framework}

\subsection{Centroid-Space Relationship}


Let $E(\cdot)$ denote a pretrained $D$-dimensional speaker encoder. For the $n$-th utterance $u_{i,n}$ of speaker $s_i$ in the training dataset, we define the utterance embedding and the speaker centroid as
\begin{equation}
\mathbf{e}_{s_i,n}
=
{E(u_{i,n})}/
{\left\|E(u_{i,n})\right\|_2},
\qquad
\mathbf{c}_{s_i}
=
\frac{1}{N_i}
\sum_{n=1}^{N_i}
\mathbf{e}_{s_i,n},
\label{eq:speaker_centroid}
\end{equation}
where $\left\|\cdot\right\|_2$ denotes the $\ell_2$ norm, and $N_i$ is the number of utterances used to construct the centroid.
For simplicity, we omit the sample index $n$ when no particular sample is specified.


For a fixed pair of speakers $(s_1,s_2)$, we similarly obtain the mixture embedding $\mathbf{e}_{\mathrm{mix},n}$ and can define
$\mathbf{c}_{\mathrm{mix}}$ by averaging embeddings constructed using different utterance pairs.
Through the investigation detailed below, we derive the following relationship:
\begin{equation}
\mathbf{c}_{\mathrm{mix}}
\approx
\frac{1}{2}
\left(
\mathbf{c}_{s_1}+\mathbf{c}_{s_2}
\right)
\quad\Longrightarrow\quad
\mathbf{c}_{s_2}
\approx
2\mathbf{c}_{\mathrm{mix}}-\mathbf{c}_{s_1}.
\label{eq:centroid_relationship}
\end{equation}

\begin{figure}[t]
    \centering
    \begin{minipage}[t]{0.495\columnwidth}
        \centering
        \includegraphics[width=\linewidth]{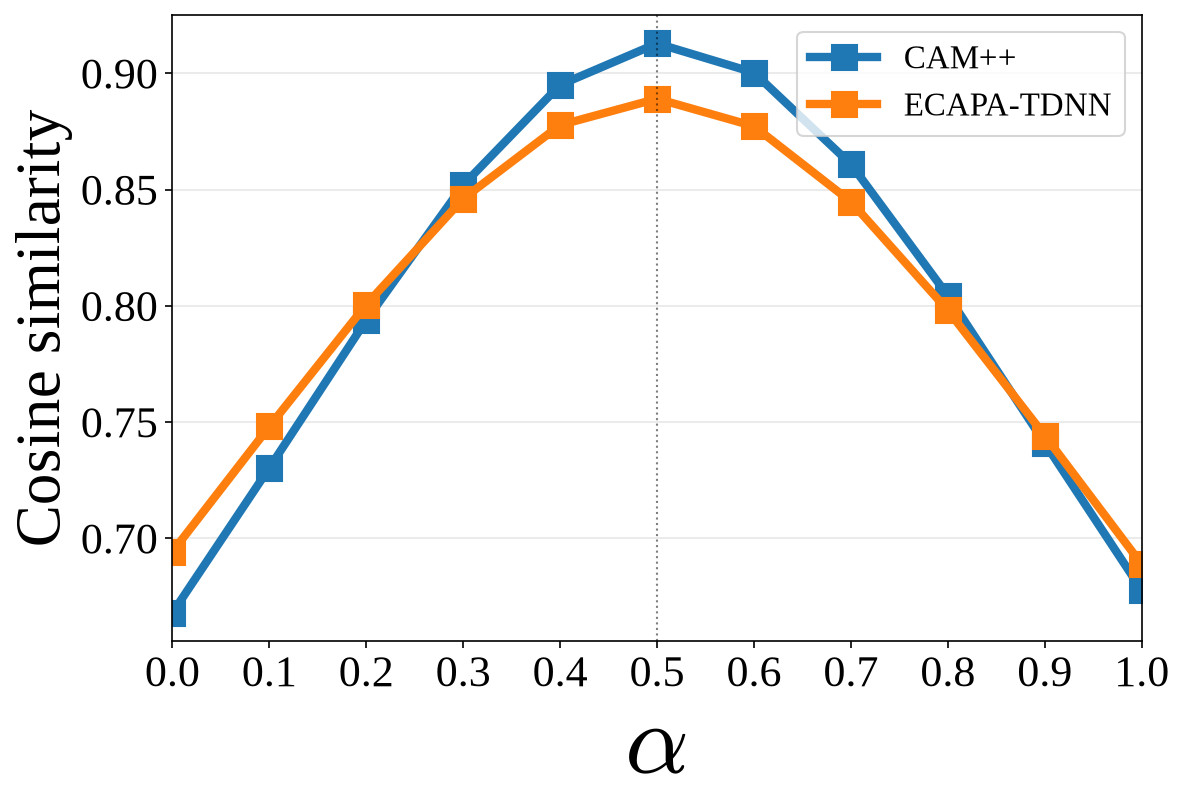}
        \vspace{-1mm}
        \centerline{(a) Interpolation weight}
    \end{minipage}
    \hfill
    \begin{minipage}[t]{0.495\columnwidth}
        \centering
        \includegraphics[width=\linewidth]{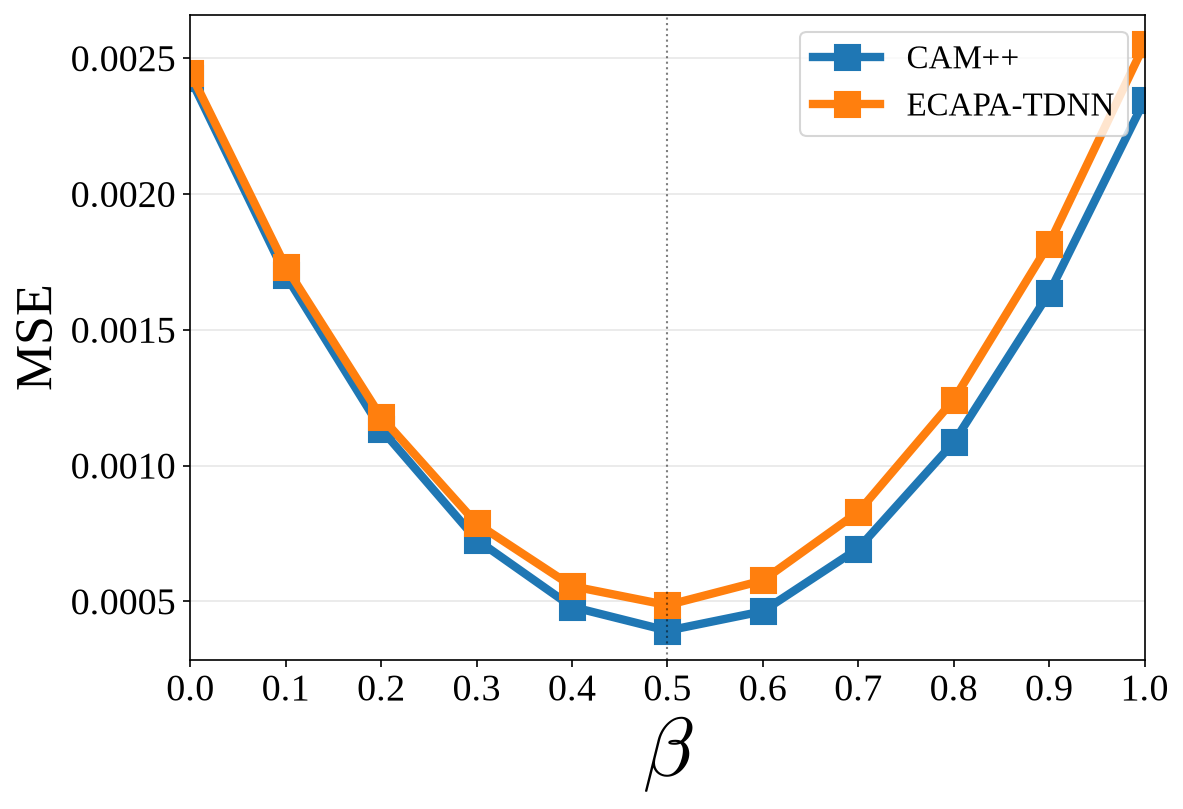}
        \vspace{-1mm}
        \centerline{(b) Centroid scaling}
    \end{minipage}
    \caption{Validation of the centroid-space relationship at $N=100$.}
        
    \label{fig:alpha_beta}
\end{figure}

\begin{figure}[t]
    \centering
    \includegraphics[width=0.9\columnwidth]{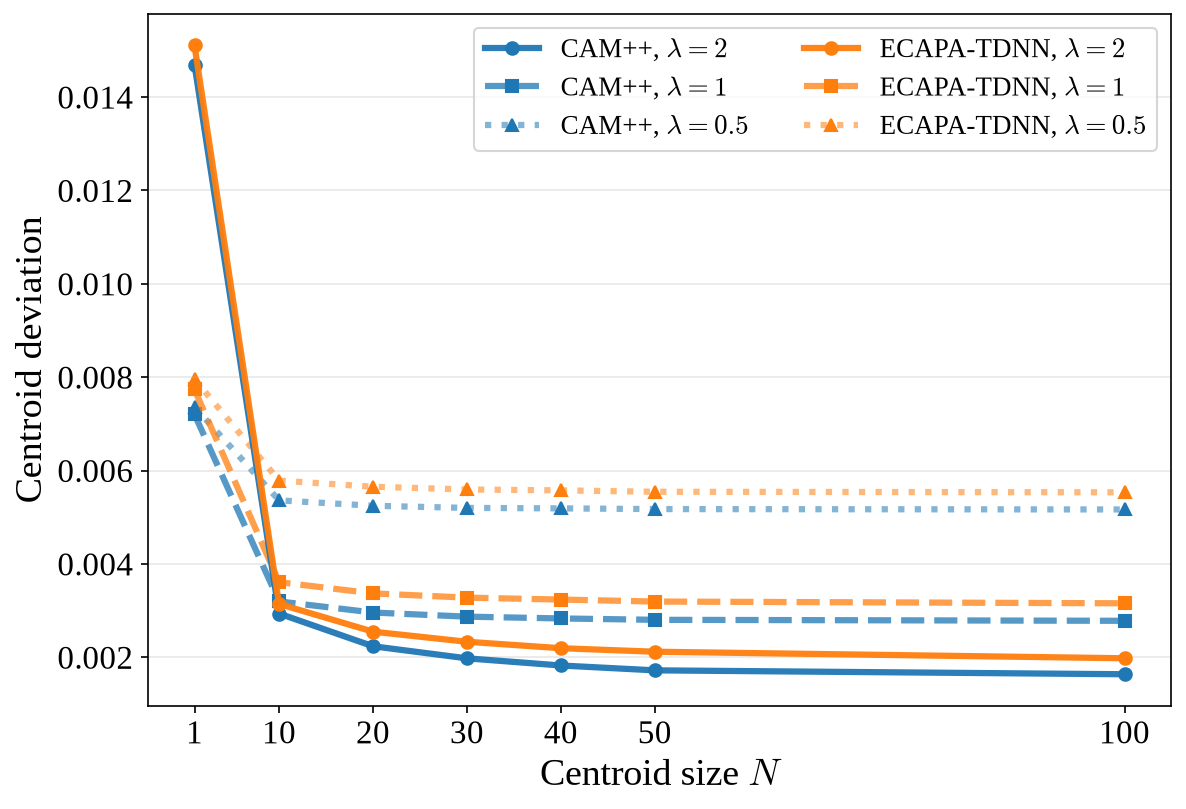}
    \vspace{-1em}
    \caption{Effect of centroid size on centroid deviation.
    Deviation is shown for different scaling factors $\lambda$ as the centroid size $N$ increases.}
    \label{fig:lambda}
\end{figure}

\begin{figure}[t]
    \centering
    \includegraphics[width=0.95\columnwidth]{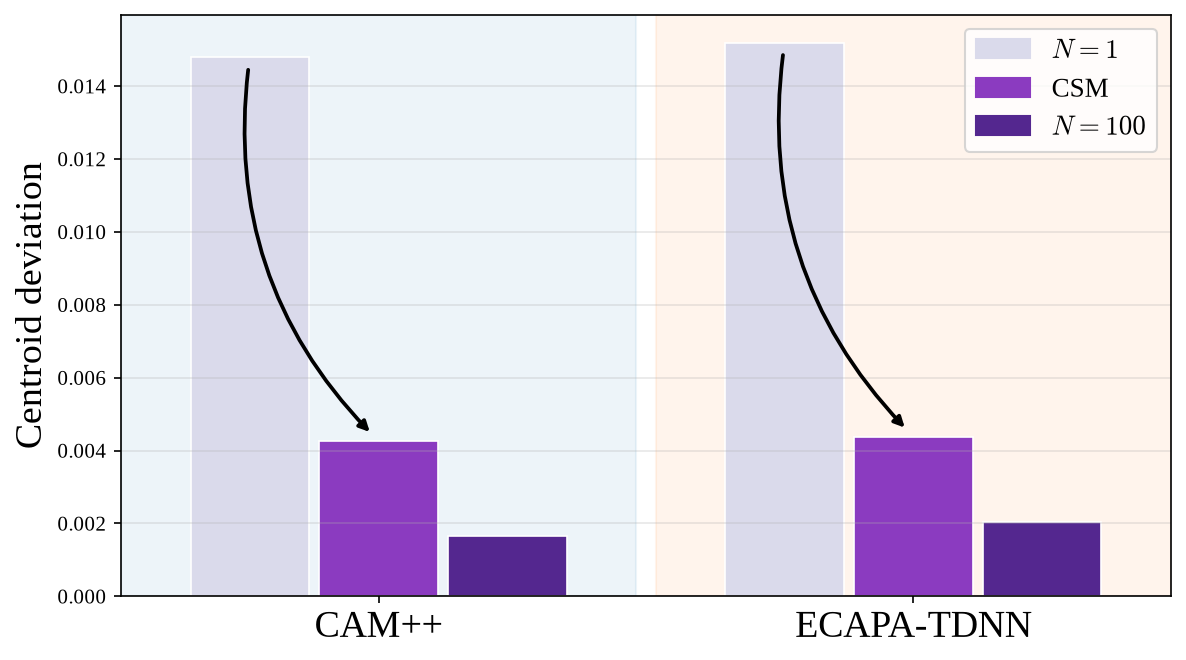}
    \vspace{-1em}
    \caption{Effect of CSM on centroid deviation.}
    \label{fig:bar}
\end{figure}

We investigated this centroid-space relationship using fixed 3-s speech and mixture segments from 30 randomly selected speaker pairs, with both CAM++~\cite{campp} and ECAPA-TDNN~\cite{ecapa}. We denote by $\mathbf{c}^{(N)}$ a centroid constructed by averaging $N$ segment embeddings. Here, we refer to $ N$ as the centroid size.
We first examined the directional relationship by varying $\alpha$ and measuring the cosine similarity between $\mathbf{c}_{\mathrm{mix}}^{(N)}$ and $\alpha \mathbf{c}_{s_1}^{(N)} + (1-\alpha)\mathbf{c}_{s_2}^{(N)}$. 
Since cosine similarity captures direction but not magnitude, and the averaged centroids are not re-normalized, we separately examined the scale by varying $\beta$ and measuring the MSE between $\mathbf{c}_{\mathrm{mix}}^{(N)}$ and $\beta(\mathbf{c}_{s_1}^{(N)} + \mathbf{c}_{s_2}^{(N)})$.
\noindent At $N=100$, Fig.~\ref{fig:alpha_beta} (a) shows that the cosine similarity is maximized
at $\alpha=0.5$, while Fig.~\ref{fig:alpha_beta} (b) shows that the MSE is minimized
at $\beta=0.5$ for both encoders. Together, these results
characterize the midpoint structure in centroid space as described in Eq.~\eqref{eq:centroid_relationship}.

To examine the effect of centroid size, we varied $N$ and measured the resulting centroid deviation. We define centroid deviation as the MSE between
$\lambda\mathbf{c}_{\mathrm{mix}}^{(N)}-\mathbf{c}_{s_1}^{(N)}$
and the oracle interferer centroid $\mathbf{c}_{s_2}$, where the oracle interferer centroid $\mathbf{c}_{s_2}$ is constructed by averaging every available full-length utterance embedding.
As shown in Fig.~\ref{fig:lambda}, the benefit of $\lambda=2$ becomes more evident as $N$ increases, while the deviation decreases and gradually saturates for both encoders. In contrast, the deviation remains substantially higher at $N=1$, indicating that the centroid-space relationship is unreliable when applied directly to individual input segments.
Since practical TSE inference typically provides only a single mixture and a single target enrollment, corresponding to $N=1$, we introduce a Centroid-Space Mapper (CSM) to recover the required centroid-space representations and estimate interferer centroids with low deviation.

\subsection{Centroid-Space Mapper}

The CSM $M_{\phi}(\cdot)$ estimates a centroid-space representation from a single input segment.
Let $x$ denote an input segment, which can be either a single-speaker signal $x_{s_i}$ or a two-speaker mixture $x_{\mathrm{mix}}$, and let $F(x)$ denote the corresponding frame-level features extracted from the frozen speaker encoder $E$ (e.g., CAM++ or ECAPA-TDNN).
Inspired by~\cite{centroid_tse}, we implement CSM with a depthwise convolutional layer, multi-head self-attention
(MHSA), and attentive statistics pooling (ASP). The CSM output $\hat{\mathbf{c}}$ denotes the centroid-space representation:
\begin{equation}
\hat{\mathbf{c}} = M_{\phi}(F(x)),
\qquad
\hat{\mathbf{c}} \in \mathbb{R}^{D}.
\end{equation}
The CSM parameters $\phi$ are shared between the enrollment and mixture branches. We train the mapper on both single-speaker signals and two-speaker mixtures. Consistent with Eq.~\eqref{eq:centroid_relationship}, the
mixture target is defined as the sum of the two speaker centroids,
which approximates twice the mixture centroid. The supervision target $\mathbf{y}$ is
\begin{equation}
\mathbf{y} =
\begin{cases}
\mathbf{c}_{s_i}, & x=x_{s_i},\\
\mathbf{c}_{s_1}+\mathbf{c}_{s_2} \approx 2\mathbf{c}_{\mathrm{mix}}, & x=x_{\mathrm{mix}},
\end{cases}
\label{eq:centroid_mapper_target}
\end{equation}
and the mapper is trained to minimize the squared $\ell_2$ error between the estimated centroid-space representation $\hat{\mathbf c}$ and the corresponding supervision target $\mathbf{y}$:

\begin{equation}
\mathcal{L}_{\mathrm{CSM}}
=
\left\|
\hat{\mathbf{c}}-\mathbf{y}
\right\|_2^2.
\label{eq:centroid_mapper_loss}
\end{equation}


At inference, CSM estimates the target centroid from the enrollment, while its mixture output approximates twice the corresponding mixture centroid:
\begin{equation}
\hat{\mathbf{c}}_{s_1}
=
M_{\phi}(F(r_{s_1})),
\qquad
2\hat{\mathbf{c}}_{\mathrm{mix}}
=
M_{\phi}(F(x_{\mathrm{mix}})).
\end{equation}

\noindent Using the centroid-space relationship in Eq.~\eqref{eq:centroid_relationship}, the interferer centroid representation is then estimated from these outputs as
\begin{equation}
\hat{\mathbf{c}}_{s_2}
=
2\hat{\mathbf{c}}_{\mathrm{mix}}
-
\hat{\mathbf{c}}_{s_1}.
\label{eq:estimated_interferer_centroid}
\end{equation}

\noindent To determine whether CSM recovers this centroid behavior from a single segment, we compare the deviation of $\hat{\mathbf{c}}_{s_2}$ from the oracle interferer centroid
$\mathbf{c}_{s_2}$ with the $N=1$ and $N=100$ references in Fig.~\ref{fig:bar}. For both encoders, CSM markedly reduces the centroid deviation relative to $N=1$ and approaches the $N=100$ reference, despite using only a single mixture and target enrollment. These results show that CSM can recover the interferer centroid
representation, which is used as a negative clue in the next section.

\begin{figure}[t!]
    \centering
    \includegraphics[width=\columnwidth]{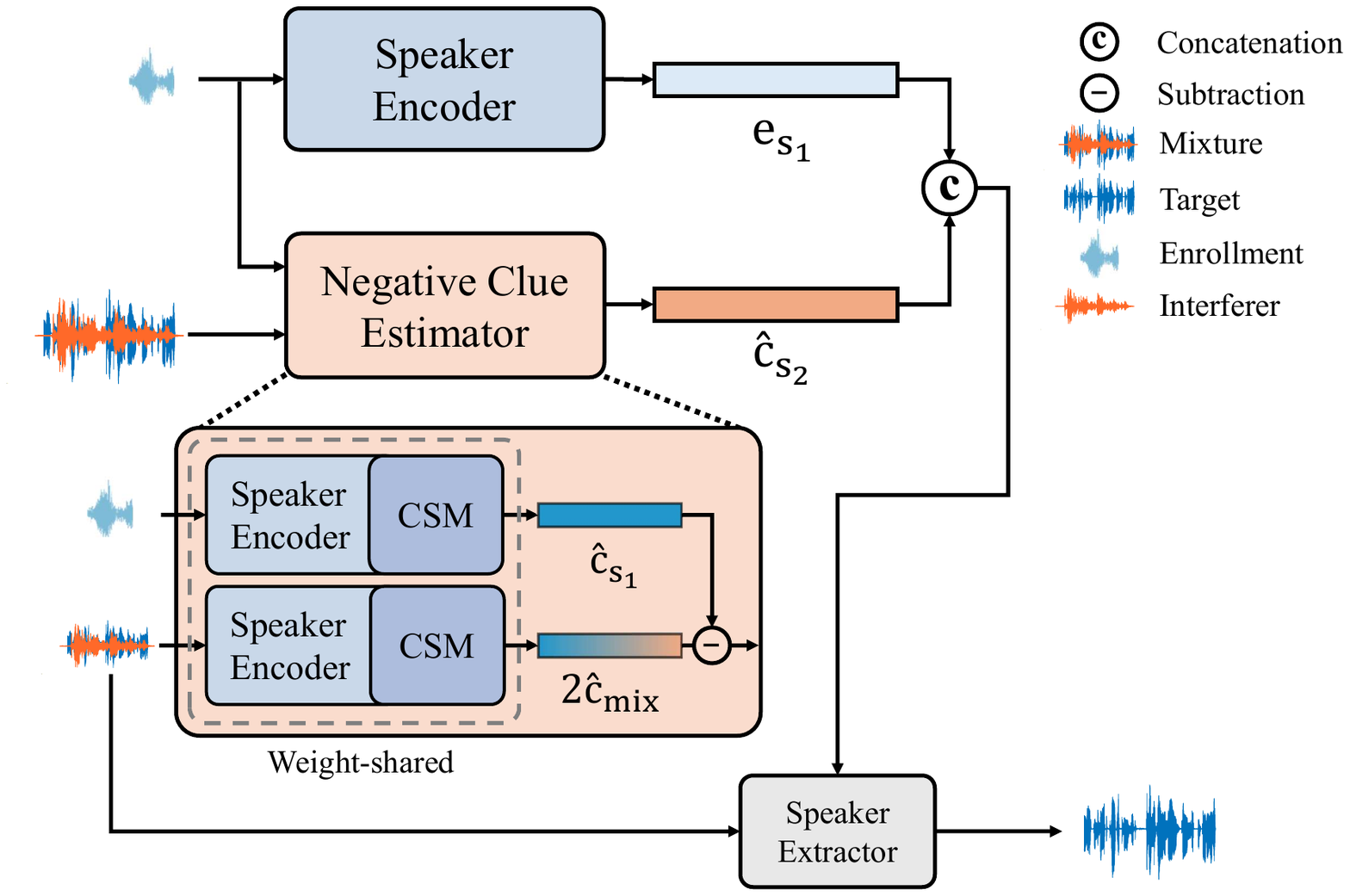}
    \vspace{-2em}
    \caption{Overview of the NECTAR inference pipeline.}
    \label{fig:overview}
\end{figure}

\subsection{Negative Clue for Target Speaker Extraction}

Given a single-channel two-speaker mixture
\begin{equation}
x_{\mathrm{mix}}=x_{s_1}+x_{s_2},
\end{equation}
our goal is to extract the target speaker $s_1$ using a target enrollment utterance.
A conventional embedding-based TSE formulation
conditions the extractor on the target speaker embedding.
The positive clue $\mathbf{P}$ is obtained from the given target enrollment utterance $r_{s_1}$, and we use the estimated interferer centroid representation from Eq.~\eqref{eq:estimated_interferer_centroid} as a negative clue $\mathbf{N}$:

\begin{equation}
\mathbf{P}
=
\mathbf{e}_{s_1}
=
E(r_{s_1})/\left\|E(r_{s_1})\right\|_2,
\qquad
\mathbf{N}
=
\hat{\mathbf{{c}}}_{s_2}
\end{equation}

\noindent Then the positive clue and negative clue are concatenated and provided to the TSE extractor:
\begin{equation}
\hat{x}_{s_1}
=
G_{\theta}
\left(
x_{\mathrm{mix}},
[
\mathbf{P};
\mathbf{N}
]
\right),
\label{eq:negative_clue_tse}
\end{equation}
where $G_{\theta}$ denotes the TSE extractor with parameters $\theta$, and $[\cdot;\cdot]$ denotes concatenation, yielding a $2D$-dimensional conditioning vector. The negative clue complements the target-speaker embedding by providing information about the speaker to be suppressed. As illustrated in Fig.~\ref{fig:overview}, NECTAR derives this negative clue from the mixture and target enrollment without requiring an interferer enrollment at inference.


\begin{table}[t]
\centering
\caption{
Comparison with TSE systems trained on LibriSpeech
train-clean-100 and evaluated on Libri2Mix.
Models marked with $^\dagger$ were trained at 8 kHz.}
\label{tab:benchmark}
\footnotesize
\setlength{\tabcolsep}{1pt}
\renewcommand{\arraystretch}{0.95}

\begin{tabular}{@{}lccc@{}}
\toprule
System
& SI-SDRi (dB)$^\uparrow$
& Acc. (\%)$^\uparrow$
& Venue \\
\midrule


sDPCCN~\cite{dpccn}
& 11.61 & -- & ICASSP'22 \\

TD-SpeakerBeam + PL + PF$^{\dagger}$~\cite{target_confusion}
& 13.88 & -- & Interspeech'22 \\

MC-SpEx$^{\dagger}$~\cite{mc_spex}
& 14.61 & -- & Interspeech'23 \\

X-T-TasNet~\cite{centroid_tse}
& 13.48 & 95.3 & Interspeech'24 \\

SSL-TD-SpeakerBeam~\cite{ssl_tse}
& 14.65 & 97.0 & ICASSP'24 \\

SHuBERT-BSRNN~\cite{selective_hubert}
& 15.39 & -- & SPL'24 \\

Multi-Level BSRNN~\cite{multilevel_tse}
& 15.91 & 97.0 & ICASSP'25 \\

\midrule

\textbf{NECTAR (Ours)}
& \textbf{16.18}
& \textbf{97.1}
& -- \\

\bottomrule
\end{tabular}
\end{table}


\section{Experimental Setup}
\label{sec:experimental_setup}

\subsection{Training}
\noindent \textbf{Datasets} We trained the TSE backbone on LibriSpeech~\cite{librispeech} train-clean-100 with dynamic speaker mixing at a signal-to-interference ratio (SIR) uniformly sampled from $[-5,5]$ dB.
Target and interferer enrollments were sampled from different utterances
of the respective speakers. For CSM training, we used LibriSpeech
train-clean-100 and Libri2Mix~\cite{librimix} train-100. All input signals were sampled at 16 kHz and segmented or zero-padded to 3 s. Oracle speaker centroids were constructed from all available utterances according to Eq.~\eqref{eq:speaker_centroid} and used as CSM supervision. Oracle mixture centroids were constructed by averaging embeddings from 100 mixtures of the same speaker pair and used as oracle references. The extractor was trained with an interferer enrollment embedding as the negative clue, while the proposed centroid-based clue was used at inference. 
We trained the extractor and CSM separately to isolate the effectiveness of the estimated negative clue and avoid adaptation to CSM-specific estimation errors. For the noisy-condition experiment, we used the same training configuration with the Libri2Mix \texttt{mix\_both} training set.

\vspace{0.2cm}
\noindent\textbf{Models}
We used BSRNN~\cite{bsrnn} as the TSE backbone and a frozen pretrained
CAM++~\cite{campp} as the speaker encoder, where $D=192$. We processed the input audio using a short-time Fourier transform (STFT) with a 32-ms Hann window and an 8-ms hop size.
The extractor was trained for 150 epochs with the negative SI-SDR loss and the Adam optimizer~\cite{adam}, using a batch size of 16 and a learning rate of $10^{-3}$ with exponential decay. The CSM used a depthwise convolution with kernel size 3, followed by a single 512-dimensional MHSA layer with 8 heads and ASP. It was trained for 50 epochs using AdamW~\cite{adamw} with a batch size of 128 and a learning rate of $4\times10^{-5}$ with exponential decay. All training was done using a single NVIDIA B200 GPU. Notably, training the CSM took about 1 hour, indicating a modest additional training cost.

\subsection{Evaluation}
We evaluated our models on the Libri2Mix \texttt{mix\_clean} test set and VCTK-2Mix\footnote{\url{https://github.com/JorisCos/VCTK-2Mix}}.
For VCTK-2Mix, speakers \texttt{p280} and \texttt{p315} were excluded due to known issues in VCTK 0.92~\cite{vctk}. We additionally evaluated the noisy-condition model on the Libri2Mix
\texttt{mix\_both} test set.
We report the scale-invariant signal-to-distortion ratio improvement (SI-SDRi)~\cite{sisdri}, perceptual evaluation of speech quality (PESQ)~\cite{pesq}, extended short-time objective intelligibility (ESTOI)~\cite{estoi}, and extraction accuracy (Acc.)~\cite{multilevel_tse}, defined as the percentage of samples with an SI-SDRi greater than 1~dB.


\begin{table}[t]
\centering
\caption{Auxiliary clue comparison on Libri2Mix
test set. The target enrollment embedding is used
as the positive clue for all methods.}
\label{tab:main}
\footnotesize
\setlength{\tabcolsep}{1pt}
\renewcommand{\arraystretch}{0.95}

\begin{tabular}{@{}llcccc@{}}
\toprule
Method
& Aux. clue 
& SI-SDRi (dB)$^\uparrow$
& PESQ$^\uparrow$
& ESTOI$^\uparrow$
& Acc. (\%)$^\uparrow$ \\
\midrule
Mixture & -- & -- & 1.15 & 0.54 & --\\
\midrule
Baseline
& --
& 14.81 & 2.79 & 0.86 & 95.4 \\

Mixture Clue
& $\mathbf{e}_{\mathrm{mix}}$
& 15.34 & 2.80 & 0.87 & 96.6 \\

Embedding Approx.
& $2\mathbf{e}_{\mathrm{mix}}-\mathbf{e}_{s_1}$
& 2.36 & 2.31 & 0.65 & 70.6 \\

\textbf{NECTAR (Ours)}
& $2\hat{\mathbf{c}}_{\mathrm{mix}}-\hat{\mathbf{c}}_{s_1}$
& \textbf{16.18}
& \textbf{2.89}
& \textbf{0.88}
& \textbf{97.1} \\

\midrule
\multicolumn{6}{l}{\textit{Oracle references}} \\

Centroid Approx.
& $2\mathbf{c}_{\mathrm{mix}}-\mathbf{c}_{s_{1}}$
& 16.62 & 2.94 & 0.89 & 98.1 \\

Oracle Centroid
& $\mathbf{c}_{s_{2}}$
& 17.44 & 2.98 & 0.91 & 99.8 \\

Oracle Utterance
& $\mathbf{e}_{s_2}^{\star}$
& 17.45 & 2.98 & 0.91 & 99.8 \\

\bottomrule
\end{tabular}
\end{table}


\section{Results}
\label{sec:Result_Analysis}

\subsection{TSE Performance}

Table~\ref{tab:benchmark} compares NECTAR with existing TSE systems
trained on LibriSpeech train-clean-100 and evaluated on Libri2Mix. NECTAR achieves the highest SI-SDRi of 16.18~dB, outperforming
Multi-Level BSRNN~\cite{multilevel_tse} at 15.91~dB, while also
achieving the highest extraction accuracy of 97.1\%.

Table~\ref{tab:main} compares different auxiliary clues for TSE on the Libri2Mix test set. We isolated the effect of the negative clue by using the target enrollment embedding as the positive clue in all methods.
\textbf{Baseline} uses only the target enrollment, while \textbf{Mixture Clue} additionally incorporates the mixture embedding
$\mathbf{e}_{\mathrm{mix}}$. NECTAR improves SI-SDRi by 1.37~dB over the Baseline and also outperforms Mixture Clue, suggesting that the derived interferer-related negative clue provides more effective conditioning than merely incorporating the mixture embedding.

The embedding approximation, which directly applies the centroid-space relationship to single embeddings, i.e., using $2\mathbf{e}_{\mathrm{mix}}-\mathbf{e}_{s_1}$ as a negative clue, severely degrades SI-SDRi to 2.36~dB.
In contrast, the oracle centroid approximation
$2\mathbf{c}_{\mathrm{mix}}-\mathbf{c}_{s_1}$ reaches 16.62~dB,
indicating that the midpoint relationship holds at the centroid level but not for individual utterance embeddings.
Using the oracle interferer centroid $\mathbf{c}_{s_2}$ further improves SI-SDRi to 17.44~dB, while the oracle utterance embedding $\mathbf{e}_{s_2}^{\star}$, extracted from the clean interferer segment corresponding to the observed mixture, achieves 17.45~dB.
These oracle results indicate both the potential benefit of providing interferer information and the validity of estimating the negative clue in the centroid space. 

On the out-of-domain VCTK-2Mix evaluation (Table~\ref{tab:vctk_ood}), NECTAR improves SI-SDRi from 12.90 to 14.67~dB without fine-tuning. It also yields consistent gains in PESQ, ESTOI, and extraction accuracy. These results indicate that the proposed negative clue generalizes well to unseen speakers and recording conditions. Moreover, as shown in Table~\ref{tab:noisy}, NECTAR consistently outperforms the Baseline under the Libri2Mix \texttt{mix\_both} condition across all metrics. This result further shows that the proposed negative clue remains effective even when the observed mixture contains background noise.


\begin{table}[t]
\centering
\caption{Out-of-domain performance on VCTK-2Mix.}
\label{tab:vctk_ood}
\footnotesize
\setlength{\tabcolsep}{4pt}
\renewcommand{\arraystretch}{0.95}

\begin{tabular}{lcccc}
\toprule

Method 
& SI-SDRi (dB)$^\uparrow$ 
& PESQ$^\uparrow$ 
& ESTOI$^\uparrow$ 
& Acc. (\%)$^\uparrow$\\
\midrule
Mixture & -- & 1.21 & 0.51 &  --\\
\midrule
Baseline
& 12.90 & 2.68 & 0.77 & 92.0\\

\textbf{NECTAR (Ours)}

& \textbf{14.67} & \textbf{2.80} & \textbf{0.80} & \textbf{94.8}\\

\bottomrule
\end{tabular}
\end{table}

\begin{table}[t]
\centering
\caption{Performance on the Libri2Mix \texttt{mix\_both} test set.}
\vspace{0.5em}
\label{tab:noisy}
\footnotesize
\setlength{\tabcolsep}{4pt}
\renewcommand{\arraystretch}{0.95}

\begin{tabular}{lcccc}
\toprule

Method 
& SI-SDRi (dB)$^\uparrow$ 
& PESQ$^\uparrow$ 
& ESTOI$^\uparrow$ 
& Acc. (\%)$^\uparrow$\\
\midrule
Mixture & -- & 1.08 & 0.40 &  --\\
\midrule
Baseline
& 11.38 & 1.85 & 0.73 & 94.3\\ 

\textbf{NECTAR (Ours)}

 & \textbf{12.02} & \textbf{1.90} & \textbf{0.74} & \textbf{95.8}\\

\bottomrule
\end{tabular}
\end{table}

\subsection{Does the centroid-space relationship hold for 3 speakers?}

\begin{figure}[!t]
    \centering
    \includegraphics[width=0.8\columnwidth]{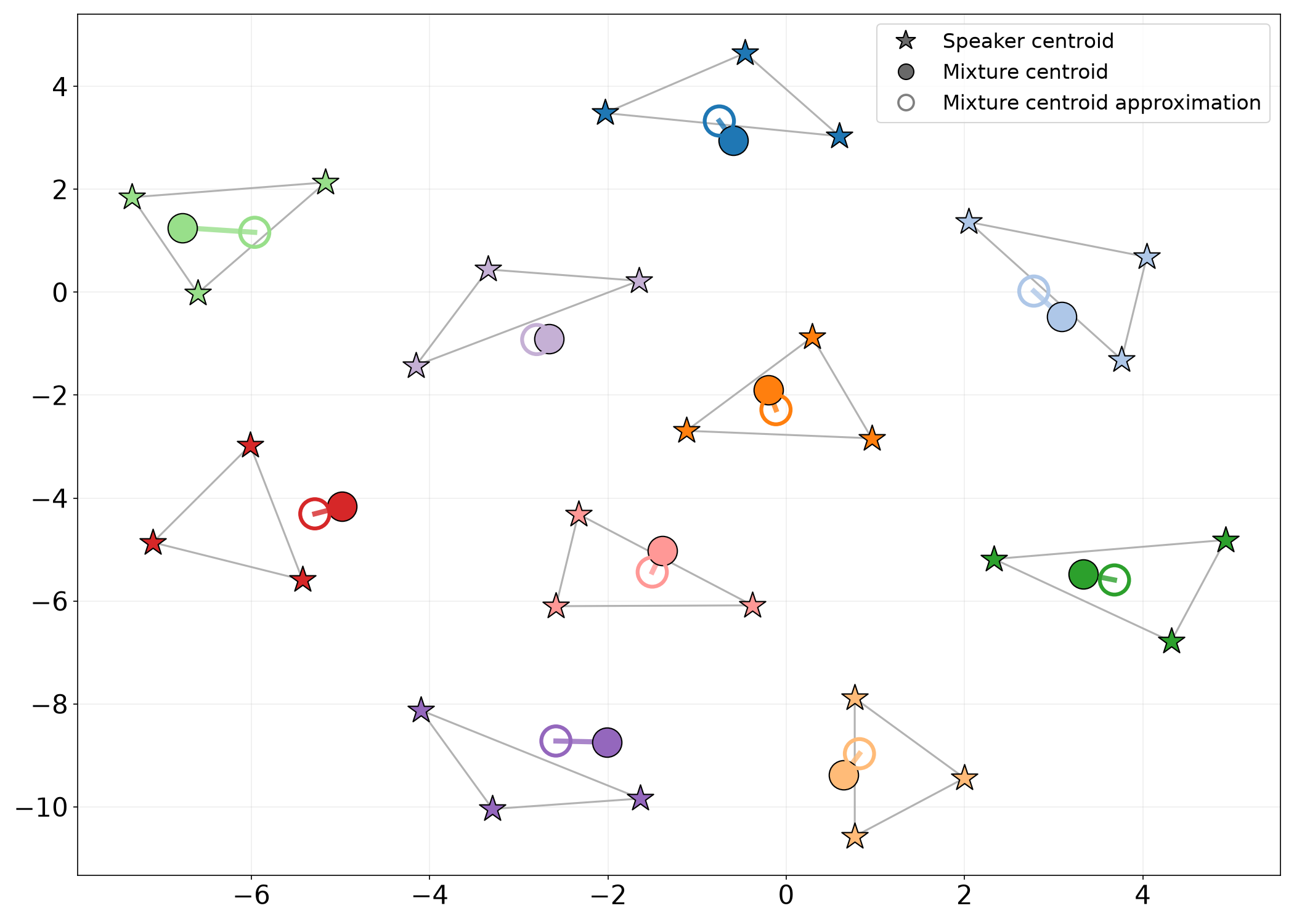}
    \vspace{-1em}
    \caption{3-speaker centroid-space relationships in t-SNE. The mean cosine similarity is 0.89 and the
    MSE is $4\times10^{-4}$.}    
    \label{fig:3mix}
\end{figure}

Furthermore, we explored the potential extension of the centroid-space relationship to three-speaker mixtures. The analysis used 10 randomly selected speaker triplets comprising 30 distinct speakers. 
Using fixed 3-s segments, all speaker and mixture centroids were constructed with $N=100$.
The mixture centroid $\mathbf{c}_{\mathrm{mix}}^{(N)}$ was compared with the equal-weight average of the three speaker centroids.
Figure~\ref{fig:3mix} visualizes the speaker centroids, the mixture
centroid, and the equal-weight centroid approximation.
The approximation achieves a mean cosine similarity of 0.89 and an MSE of $4\times10^{-4}$, indicating a similar geometric structure in the three-speaker setting.


\section{Conclusion}
\label{sec:conclusion}

This work shows that interferer information for TSE can be derived from the observed mixture and the given enrollment through centroid-space structure. 
Our analysis revealed an approximate midpoint relationship between the mixture centroid and its constituent speaker centroids, and CSM enabled this centroid-space relationship to be exploited from individual input segments.
NECTAR achieved 16.18~dB SI-SDRi on Libri2Mix. The performance gains extend to noisy conditions and to the out-of-domain VCTK-2Mix evaluation without fine-tuning. 
Overall, these results demonstrate that incorporating derived interferer information with the target representation provides effective, generalizable conditioning for TSE.

\section{Limitations}

Although our analysis provides empirical evidence for the midpoint relationship in both CAM++ and ECAPA-TDNN, a theoretical explanation of the midpoint relationship remains underexplored. Moreover, we only explore two-speaker TSE, though the three-speaker centroid analysis suggests potential applicability. Extending NECTAR to mixtures with more speakers and to settings where the primary interfering source is non-speech remains future work.

\vfill\pagebreak
\section{Acknowledgment}
This work was supported by the National Research Foundation of Korea (NRF) grant (No. RS-2024-00337945), the STEAM research grant (No. RS-2024-00464269) funded by the Ministry of Science and ICT of Korea government (MSIT), Hyundai Motor Company and Kia (No. G01250273), the BK21 FOUR program through the NRF grant funded by the Ministry of Education of Korea government (MOE), Artificial intelligence industrial convergence cluster development project funded by the Ministry of Science and ICT (MSIT, Korea)\&Gwangju Metropolitan City, and Industrial Technology Innovation R\&D program of MOTIE/KEIT (No. RS-2025-24533624).


\bibliographystyle{IEEEbib}
{\footnotesize
\bibliography{strings,refs}
}
\clearpage

\end{document}